# Anti-damping spin transfer torque through epitaxial Nickel oxide

Takahiro Moriyama[1], So Takei[2], Masaki Nagata[1], Yoko Yoshimura[1], Noriko Matsuzaki[1], Takahito Terashima[3], Yaroslav Tserkovnyak[2], and Teruo Ono[1]

1. *Institute for Chemical Research, Kyoto University, Japan.*
2. *Department of Physics and Astronomy, University of California, Los Angeles, CA 90095, USA*
3. *Research Center for Low Temperature and Materials Sciences, Kyoto University, Japan.*

Abstract

We prepare the high quality epitaxial MgO(001)[100]/Pt(001)[100]/NiO(001)[100]/FeNi/SiO$_2$ films to investigate the spin transport in the NiO antiferromagnetic insulator. The ferromagnetic resonance measurements of the FeNi under a spin current injection from the Pt by the spin Hall effect revealed the change of the ferromagnetic resonance linewidth depending on the amount of the spin current injection. The results can be interpreted that there is an angular momentum transfer through the NiO. A high efficient angular momentum transfer we observed in the epitaxial NiO can be attributed to the well-defined orientation of the antiferromagnetic moments and the spin quantization axis of the injected spin current.

Since the theoretical predictions[1,2] followed by experimental demonstrations [3,4], spin transfer torque (STT) has been an efficient and promising technique to control magnetizations of ferromagnetic materials in modern spintronics devices. This novel technique is based on an interaction between electron spin and local magnetic moments. Namely, the angular momentum of the electron spin is transferred to and exerts a torque on the magnetization. The same interaction should be conserved in antiferromagnets (AFMs), in which there are microscopic local magnetic moments that compensate each other to exhibit no net magnetization[5,6,7,8]. As AFMs have been abandoned as an active material in spintronics in spite of their potential applications in the THz regime[9], it is of great interest to investigate the STT in AFMs.

Despite numerous theoretical works on STT in AFMs, there are only a few experiments indicating the possibilities of interactions between the electron spin and the AFM moments[10,11,12,13]. The most recent investigations of spin transport in AFM materials have raised an interesting question, whether AFMs can be transparent to the angular momentum flow[12,13]. Wang et al.[13] performed spin pumping measurements on $Y_3Fe_5O_{12}$/NiO/Pt and a spin voltage signal was detected by the inverse spin Hall effect in Pt. Spin transfer in a metallic AFM system with a Pt/IrMn/FeCoB heterostructure has also been investigated by our group by means of the spin-torque ferromagnetic resonance (ST-FMR) [12]. In the ST-FMR measurements, it was found, by investigating the change in linewidth, that the spin current injected from the Pt can give a spin torque on the FeCoB via the IrMn. It is quite remarkable that indications of spin transfer through AFMs have been confirmed by these two different techniques.

Those observations mostly draw only macroscopic interpretation of the interaction between the AFM moments and the spin current because the Neel vectors are most likely randomly oriented with respect to the spin current due to polycrystallinity of the AFMs. In order to understand the microscopics of the interaction between AFM moments and the spin current, it is more desirable to investigate a clean and ordered structure that possesses well-defined AFM moments.

In this letter, we prepared MgO(001) substrate/Pt/NiO/FeNi/SiO$_2$ multilayers, in which the films are epitaxially grown until the NiO layer, and performed a ST-FMR measurement to quantify the anti-damping spin torque transported between the Pt and the FeNi through the NiO layer. In order to investigate the spin torque in NiO, we created and injected a pure spin current by the spin Hall effect in Pt. The schematic layer structure of the injection scheme is shown in Fig. 1(b). The electron flowing in the Pt layer experiences a spin dependent scattering due to spin-orbit interaction, resulting in the opposite flow of spin polarized electrons orthogonal to the electron flow[14]. This spin current is a so-called pure spin current, which does not involve charge current flow. This spin current induces a spin accumulation at the Pt/NiO interface and exerts a spin torque on the NiO magnetic moments in a similar manner to the Pt/ferromagnetic insulator case[15]. We investigate the magnetic damping of the FeNi grown on top of the NiO under the pure spin current injection to understand the STT in the NiO layer. Unlike ferromagnetic materials, magnetic moments in oxide antiferromagnetic materials cannot be easily controlled by external magnetic field and they are tied with the crystalline magnetic anisotropy[16]. In order to control the AFM magnetic moments, we develop a single crystal NiO. Nickel oxide has a rock salt structure with the lattice constant of 4.2 Å[17], which is similar to the lattice constants of MgO (4.1 Å) and Pt (3.9 Å) within a 5% mismatch. The magnetic moments inhabit the nickel atoms as shown in Fig. 1(a). The magnetic moments in {111} planes are ferromagnetically coupled with its anisotropy in the <-211>, <1-21>, and <11-2> directions. The antiferromagnetism originates from the antiparallel coupling between the magnetic moments in the neighboring {111} planes by super exchange coupling. From the symmetry of these crystal planes, there are twelve possible orientations of magnetic moments, each of which makes a non-collinear angle with the [010] directions.

The layer stacks of Pt 5nm/NiO 10 nm/FeNi 3 nm/SiO$_2$ 5 nm were prepared on MgO(001) single crystal substrate (Fig. 1(b)). Pt layer was formed by magnetron sputtering on MgO(001) with a substrate temperature of 600 ºC and was then taken out to the air. After transferring the sample to another chamber for pulsed laser deposition (PLD), the sample was heated to 400 ºC and its surface crystallinity was checked by reflection high energy electron diffraction (RHEED). Fig. 1(c)(i) shows the RHEED pattern of the Pt surface in the direction of MgO(001)[100]. Clear streaks indicate the atomically smooth surface of Pt layer and the crystal orientation relative to the MgO is determined to be MgO(001)[001]/Pt(001)[001]. NiO layer was then deposited by PLD with an oxygen partial pressure of $2 \times 10^{-2}$ Pa at the substrate temperature of 400 ºC. The RHEED pattern for the surface of the NiO 10nm shown in Fig. 1(c)(ii) also shows clear streaks indicating the atomically smooth surface from which we determined the epitaxial orientation relationship between the NiO and the Pt layer to be MgO(001)[100]//Pt(001)[100]//NiO(001)[100]. The following FeNi and SiO$_2$ layers were deposited by magnetron sputtering. The samples shown in this letter are as-deposited with no unidirectional anisotropy in FeNi. Control samples of MgO(001)/Pt 5nm/SiO$_2$ 10 nm/FeNi 3 nm/SiO$_2$ 5 nm and MgO(001)/Pt 5nm /FeNi 3 nm/SiO$_2$ 5 nm, all made by magnetron sputtering, were grown epitaxially up to the Pt layer.

In order to investigate the spin transport in the NiO, we performed the ST-FMR

measurement, which is a useful method for quantifying the spin torque exerting in the system. The film is patterned into 4 ~ 10 μm wide strips attached to a coplanar waveguide facilitating both the r.f. and d.c. current injection into the strip. The long axis of the strip along which the d.c. current flows is parallel to [100] direction of the MgO substrate. This ensures the spin quantization axis of the pure spin current to be in the [010] direction, thus always making a finite angle with the anisotropic magnetic moments of the epitaxially grown NiO, and so that the spin current would exert a spin torque on the NiO magnetic moments. ST-FMR is performed by sweeping the external magnetic field at a fixed frequency of the r.f. current. Figure 2(a) shows the measurement configuration together with our coordinate system. The positive electric current is defined when it flows along the positive $x$ direction. The external positive magnetic field is applied in the sample plane and at 45º away from $x$ axis. We apply nominal r.f. power up to 14 dBm and d.c. current up to 2.5 mA to the strip. All the measurements were performed at room temperature. The expected FMR spectra with the spin injection by the spin Hall effect in such a system can be composed of symmetric and antisymmetric Lorentzians. The detailed explanation can be found elsewhere[12,18]. The obtained spectra shown in Fig. 2(b) for the Pt/NiO/FeNi/SiO$_2$ are composed of both Lorentzians and the ratio between the symmetric and antisymmetric components is found to be finite, implying that the spin torque is exerted on the FeNi layer through the NiO. In order to further characterize the spin torque in the NiO layer, we mainly focused on the change in the linewidth under the spin current injection which reflects the change in effective damping due to the anti-damping spin torque due to the spin current.

Figures 3 summarizes the results obtained by the ST-FMR measurement and compares with the control samples. The most important results in this letter on the linewidth change due to the spin current are shown in Fig.3 (g-i). Before discussing our central results, we would like to first show the basic FMR results obtained by the ST-FMR measurement. The ferromagnetic resonance for three different samples summarized in Figs. 3 (a-c) are well fitted by Kittel's formula $f = (\gamma/2\pi)\sqrt{(H_{ex} + H_a)(H_{ex} + H_a + 4\pi M_{eff})}$, where $\gamma = 1.76 \times 10^{11}$ [s$^{-1}$ T$^{-1}$] is gyromagnetic ratio, $H_{ex}$ is the external magnetic field, $H_a$ is the inplane anisotropy field, and $4\pi M_{eff}$ is the effective demagnetizing field. The fitting yields $4\pi M_{eff}$ = 0.59, 0.72, and 0.6 T with a negligible inplane anisotropy for Pt/NiO 10nm/FeNi, Pt/FeNi, and Pr/SiO$_2$/FeNi samples, respectively. We separately measured the saturation magnetization $M_S = 0.78$ T for a single layer of the FeNi by SQUID magnetometry. The discrepancy between the saturation magnetization and the $4\pi M_{eff}$ values may originate from the perpendicular anisotropy or magnetic dead layer at the interface. The Gilbert damping $\alpha$ of the FeNi is extracted from the linewidth $\Delta H$ as a function of frequency by $\Delta H = \Delta H_0 + 2\pi f \alpha/\gamma$ where $\Delta H_0$ is the inhomogeneous broadening[19]. We obtain $\alpha$ = 0.016, 0.04, and 0.009 and $\Delta H_0$= 14, 16, and 6 Oe for Pt/NiO 10nm/FeNi, Pt/FeNi, and Pr/SiO$_2$/FeNi, respectively, as the actual fittings are shown in Figs. 3(d-f). Finally, the most important observations on the change of the linewidth by the spin current are shown in Figs. 3 (g-i). It is clearly seen that the change in the linewidth is observed in the Pt/NiO/FeNi sample as well as the Pt/FeNi sample and not in the Pt/SiO$_2$/FeNi sample. As explained above, the d.c. current flowing in the Pt layer invokes the spin Hall effect and induces a spin accumulation at the Pt/NiO interface. The spin torque transferred into the ferromagnetic layer should result in either increase or decrease in the linewidth as the typical Pt/FeNi linewidth change can be seen in Fig.

3(h). With the positive spin Hall angle and the positive external field, the magnetization in FeNi feels the anti-damping torque in the positive d.c. current, leading to the decrease of the FMR linewidth[18]. On the other hand, by flipping the magnetization direction with the negative external field, the positive d.c. current results in the increase of the linewidth. As the ST-FMR measurement is only sensitive to the linewidth (i.e. damping) of the FeNi layer, the change in the linewidth in Pt/NiO/FeNi can be interpreted in the way that the angular momentum is indeed transferred through the NiO[13]. No linewidth change in the Pt/SiO$_2$/FeNi sample indicates that the angular momentum transport is completely impeded by non-magnetic SiO$_2$ layer as one expects.

The relation between the linewidth and the injected spin current is given by,

$$\delta H/I_{dc} = \frac{1}{\sqrt{2}} \frac{2\pi f}{\gamma} \frac{1}{(H_{ex} + 2\pi M_{eff})M_s t_{FM} w} \frac{\hbar}{2e} \left(\beta \frac{\theta_{SH,Pt} r_{Pt}}{t_{Pt}}\right) \quad (1)$$

where $\delta H$ is the change in linewidth, $H_{ex}$ is the external field, $\hbar$ is the Planck's constant, $e$ is the elementary charge, $w$ is the width of the strip, $t_{FM}$ is the thickness of FeNi, $\theta_{SH,Pt}$ is the spin Hall angle of the Pt, $r_{Pt}$ is the shunt current ratio of the Pt layer, and $t_{Pt}$ is the thickness. Here, we introduce $\beta$ ($0 \leq \beta \leq 1$) as the spin-transfer efficiency in the NiO layer. $\beta = 1$ when all of the injected spin current exerts a spin torque on the FeNi layer through the NiO layer (and also corresponding to the limit of zero NiO thickness), and $\beta = 0$ when it is dissipated before reaching to the FeNi layer. Using the effective spin Hall angle $\theta_{SH,Pt}$ for the Pt/FeNi interface for both Pt/FeNi and Pt/NiO interfaces and the same shunt current ratios in Pt for both Pt/NiO/FeNi and Pt/FeNi samples, we can simply compare $\delta H/I_{dc}$ values obtained from the slope of the plots in Figs. 3(g) and (h) to see how the NiO efficiently carries the spin angular momentum. By fitting with the data in Figs. 3(g) and 3(h), we obtained $\delta H/I_{dc}$ to be 1.1 ± 0.1 and 1.2 ± 0.1 Oe/mA, respectively, for Pt/NiO/FeNi and Pt/FeNi samples.

These results yield a highly efficient spin transport with $\beta = 0.82$ for the NiO layer. Assuming spin transport through NiO decays exponentially over a decay length $\lambda_s$ (see theoretical discussion below) such that $\beta \equiv e^{-t_{AFM}/\lambda_s}$ in Eq. (1), the $\beta$ value obtained above implies $\lambda_s \approx 50$nm for $t_{AFM} = 10$nm. Comparing this with a previous report[13], in which spin transport through NiO indicated a decay length of order 10nm, our $\beta$ value implies a much more efficient angular momentum transfer.

At finite temperatures, spin can be transported through NiO by the coherent Neel order dynamics as well as incoherent thermal magnons. While the incoherent contribution is expected to decay exponentially over the spin diffusion length $\lambda_{sd}$, the coherent contribution exponentially decays over the so-called healing length $\lambda_h$ (as discussed below). The experimental decay length $\lambda_s$ should then be determined by $\lambda_s \equiv \max\{\lambda_{sd}, \lambda_h\}$. A theory for spin transport through AFM insulators mediated by coherent Neel order dynamics (i.e., thermal contribution neglected) has been developed in Refs. 7 and 20. According to the above discussion, such a theory is sufficient either at temperatures well below the Neel temperature (at which thermal magnon density is negligible) or for $\lambda_{sd} \ll \lambda_h$. Specifically, Ref. 20 considers a metal|AFM|ferromagnet trilayer (akin to the current experiment), in which spin is injected into the AFM from the metal via the spin Hall effect and the spin transfer torque mechanism (modeled using spin Hall phenomenology[21]), transported through the AFM by its coherent collective excitations, and transferred into the ferromagnet via AFM/ferromagnet exchange coupling. NiO is known to have a magnetic easy-plane parallel to the (111) plane, within which a much weaker easy-axis exists with anisotropy parameter $\kappa \approx 33$ J/m$^3$ [22]. Ref. 20 considers an AFM with easy-axis anisotropy and obtains an analytical result for the FMR linewidth within linear-response in the charge current.

Applying the result for the case where the charge current direction and the FMR d.c. field makes an angle 45° (as in this experiment), the theoretical efficiency parameter $\beta_{th}$ entering Eq. (1) becomes (see appendix),

$$\beta_{th} = \frac{1}{\left[\cosh\left(\frac{t_{AFM}}{\lambda_h}\right) + \frac{1}{\eta}\sinh\left(\frac{t_{AFM}}{\lambda_h}\right)\right]^2}, \quad (2)$$

where the healing length $\lambda_h = \sqrt{A/\kappa}$ ($A$ being the NiO exchange stiffness), and $\eta = H_e M_s t_{FM}/\sqrt{A\kappa}$ ($H_e$ being the exchange coupling field for the NiO/FeNi interface). Using $H_e \sim 10^2$ Oe [23] and $A \sim 5 \times 10^{-12}$ J/m [22], we obtain $\eta \sim O(1)$, and find that $\beta \sim 0.82$ for $t_{AFM}/\lambda_h \sim 0.1$. Since $\lambda_h = \sqrt{A/\kappa} \sim 100$nm, the NiO thickness of $t_{AFM} = 10$nm is consistent with this result. The decay length $\lambda_s$ obtained above is also in reasonable agreement with $\lambda_h$ obtained here.

In summary, we performed the ST-FMR measurement on NiO and observed the change of the linewidth as a function of the spin current from the Pt due to the spin Hall effect. Our observation is interpreted that there is an angular momentum transfer from the Pt to the FeNi through the NiO. Our high efficient angular momentum transfer in the NiO comparing with the previous report on disordered NiO indicates that the orientation of the AFM moments is important, which presumably support the available theoretical model assuming the angular momentum transfer mediated by AFM spin fluctuations. Our results also assure that the NiO antiferromagnetic moments are interacting with the spin current and exciting dynamics. It would be of great interest to experimentally prove the spin torque magnetic excitation in antiferromagnet as the similar concept has already been demonstrated in ferromagnetic thin films [24]. Furthermore, a study of the linewidth at lower temperatures will effectively freeze out the thermal magnons and allow one to quantify spin transport mediated purely by the coherent Neel dynamics, and the results for the linewidth as a function of NiO thickness at these low temperatures can then be used to extract the AFM healing length [c.f. Eq. (2)]. Finally, both theoretical and experimental study of the temperature-dependence of the linewidth is still lacking but interesting, for it will shed light on how thermal magnons contribute to the AFM spin transport.

**Acknowledgements**

This work was partly supported by Grants-in-Aid for Scientific Research (S), Grant-in-Aid for Young Scientists (B) from Japan Society for the Promotion of Science. ST and YT would like to acknowledge support by FAME (an SRC STARnet center sponsored by MARCO and DARPA).

**Appendix**

In this appendix, we derive Eq. (2). Within the theoretical model of Ref. 20, the change in linewidth coming from the STT effect is[20]

$$\delta H = \frac{\omega_{res}}{\mu\gamma}\frac{\vartheta j}{b_0 S}\frac{\sin\zeta}{\left[\cosh\left(\frac{t_{AFM}}{\lambda_h}\right) + \frac{1}{\eta}\sinh\left(\frac{t_{AFM}}{\lambda_h}\right)\right]^2},$$

where $\zeta$ is the angle subtended between the FMR d.c. field and the charge current. The above result is valid to linear-order in the current density $j$ in Pt. Here, $\mu$ and $S$ are the magnetic permeability of and the total spin (in units of $\hbar$) in the FeNi, respectively; the rest of the parameters are as defined in the main text. According to spin Hall phenomenology,[21] the dissipative torque coefficient is given by

$$\vartheta = \frac{\hbar\mathcal{A}}{2|e|}\tan\theta_{Pt}$$

where $\theta_{Pt}$ is the effective spin Hall angle for Pt and $\mathcal{A}$ is the interface cross-sectional area. Then for $\theta_{Pt} \ll 1$, we obtain

$$\frac{\delta H}{I_{dc}} \approx \frac{\omega_{res}}{\mu\gamma S b_0} \frac{\mathcal{A}}{w} \left(\frac{\hbar\theta_{Pt}}{2|e|t_{Pt}}\right) \frac{\sin\zeta}{\left[\cosh\left(\frac{t_{AFM}}{\lambda_h}\right) + \frac{1}{\eta}\sinh\left(\frac{t_{AFM}}{\lambda_h}\right)\right]^2} .$$

Note that $\mu\gamma S = M_S \mathcal{A} t_{FM}/\hbar$, where $M_S$ is the saturated magnetization in units of Tesla (as given in the main text). With $b_0 \approx H_{ex} + 2\pi M_{eff}$ and $\zeta = 45°$. To be in line with the experiment, we use the same $\theta_{Pt}$ for both Pt/FeNi and Pt/NiO/FeNi cases, and arrive at the theoretical expression for the efficiency parameter $\beta_{th}$ given in Eq. (2). We note that $\beta_{th}$ does not account for the possible differences in the effective spin Hall angles and shunt current ratios between the Pt/FeNi and Pt/NiO/FeNi samples, while these differences are contained in the experimentally obtained value for $\beta$.

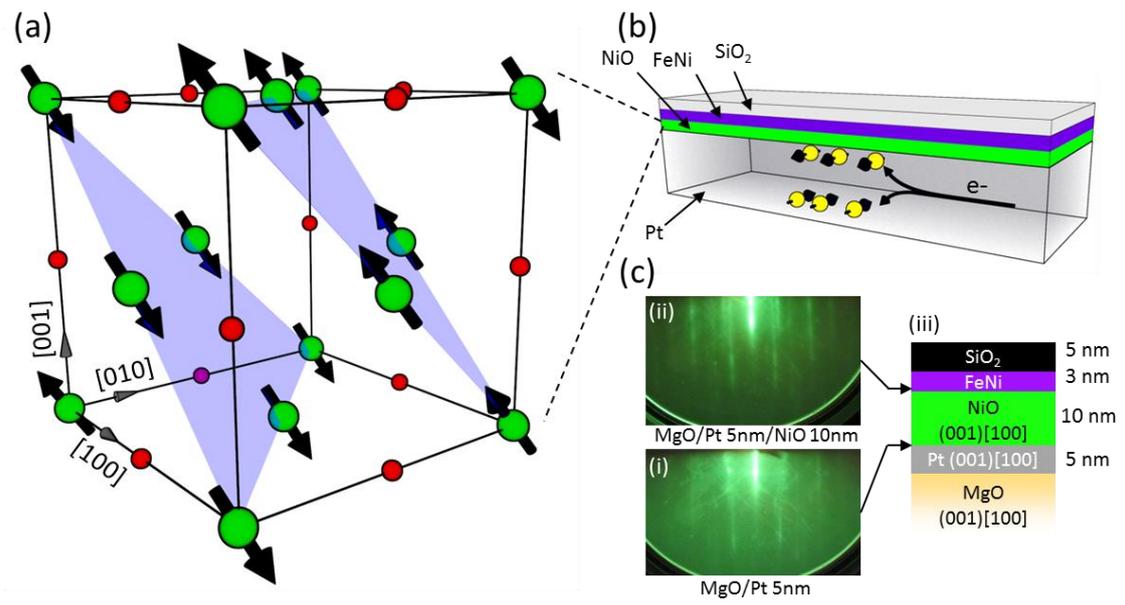

Figure 1 Moriyama et al.

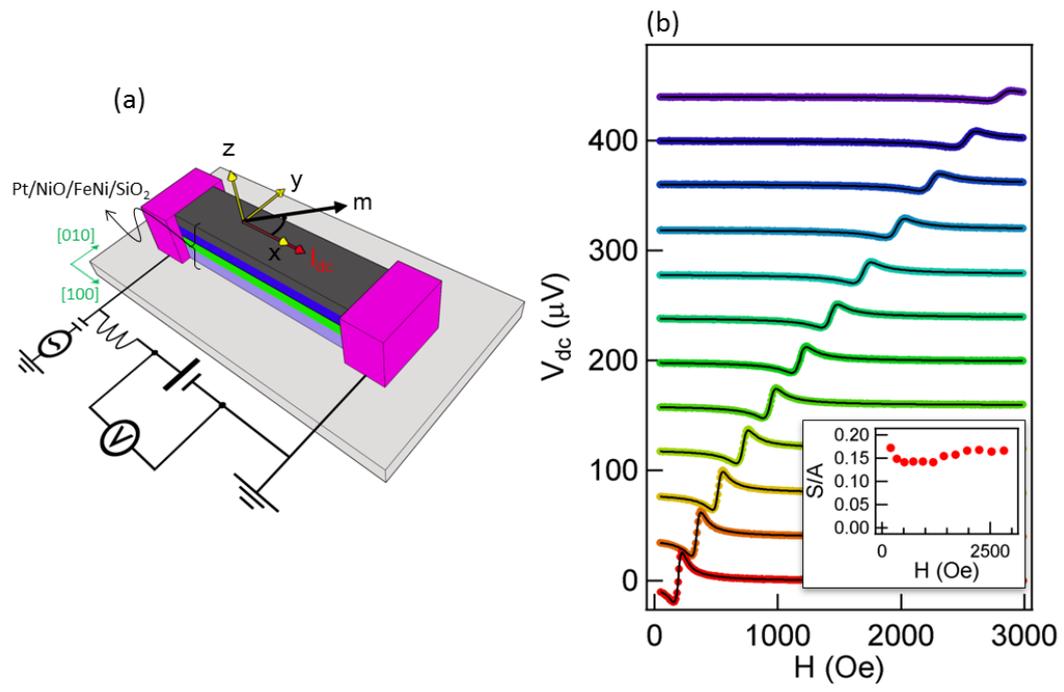

Figure 2 Moriyama et al.

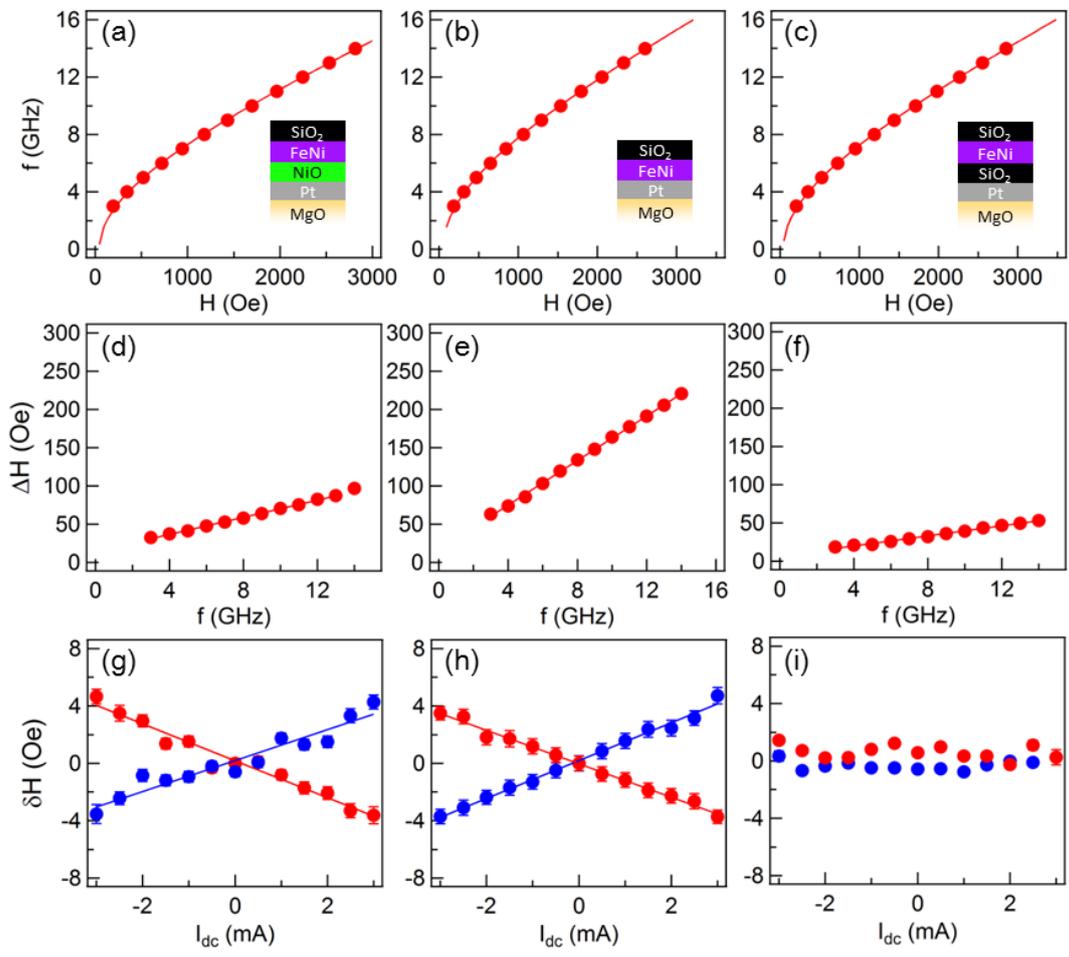

Figure 3 Moriyama et al.

Figure captions:

FIG. 1 (a) Crystal lattice of NiO. Green and red spheres indicate the Ni and O atoms, respectively. Magnetic moments reside at the Ni atoms. (b) Schematic illustration of the spin injection scheme. The bottom Pt layer invokes an injection of the pure spin current due to the spin Hall effect. (c) reflection high energy electron diffraction images for the Pt surface (i) and the NiO surface (ii). Schematic of the crystal orientation relationship is shown in (iii).

FIG. 2 (a) Measurement configuration with a ST-FMR circuitry. The strip sample is patterned so that $I_{dc}$ flows along the [100] direction of the MgO. (b) ST-FMR spectra for Pt 5nm/NiO 10nm/FeNi 3nm/SiO$_2$ 5nm. The inset shows the ratio of symmetric (S) and antisymmetric (A) Lorentzian magnitudes in the spectra.

FIG. 3 (a-b) Resonant frequency vs. field curves (d-f) The FMR spectral linewidth as a function of the frequency (g-i) The change of the linewidth as a function of $I_{dc}$ at the resonance frequency of 9GHz for Pt 5nm/NiO 10nm/FeNi 3nm/SiO$_2$ 5nm, Pt 5nm/FeNi 3nm/SiO$_2$ 5nm, and Pt 5nm/SiO$_2$ 10nm/FeNi 3nm/SiO$_2$ 5nm, respectively. The dots are the experimental data and the lines are the fitting as explained in the main text.

[1] J.C. Slonczewski, J. Magn. Magn. Mater. 159, L1 (1996)

[2] L. Berger, Phys. Rev. B 54, 9353 (1996)

[3] M. Tsoi, A. G. M. Jansen, J. Bass, W.C. Chiang, M. Seck, V. Tsoi, P. Wyder, Phys. Rev. Lett. 80 (1998) 4281.

[4] J. A. Katine, F. J. Albert, R. A. Buhrman, E. B. Myers, D. C. Ralph, Phys. Rev. Lett. 84, 3149 (2000)

[5] A. H. MacDonald and M. Tsoi, Phil. Trans. R. Soc. A 369, 3098 (2011).

[6] A. C. Swaving and R. A. Duine, Phys. Rev. B 83, 054428 (2011)

[7] S. Takei, B. I. Halperin, A. Yacoby, and Y. Tserkovnyak, Phys. Rev. B 90 094408 (2014)

[8] R. Cheng, J. Xiao, Q. Niu, and A. Brataas, Phys. Rev. Lett. 113 057601 (2014)

[9] T. Kampfrath, A. Sell, G. Klatt, A. Pashkin, S. Mährlein, T. Dekorsy, M. Wolf, M. Fiebig, A. Leitenstorfer, R. Huber, Nat. Photon. 5, 31 (2011)

[10] Z. Wei, A. Sharma, A. S. Nunez, P. M. Haney, R. A. Duine, J. Bass, and A. H. MacDonald, and M. Tsoi, Phys. Rev. Lett. 98, 116603 (2007)

[11] S. Urazhdin and N. Anthony, Phys. Rev. Lett. **99**, 046602 (2007).

[12] T. Moriyama et al., arXiv:1411.4100

[13] H. Wang, C. Du, P. C. Hammel, and F. Yang, Phys. Rev. Lett. 113, 097202 (2014)

[14] J. –P. Jan, Solid State Phys. **5, 1** (1957)

[15] Y. Kajiwara, K. Harii, S. Takahashi, J. Ohe, K. Uchida, M. Mizuguchi, H. Umezawa, H. Kawai, K. Ando, K. Takanashi, S. Maekawa, and E. Saitoh, Nature 464, 262 (2010)

[16] K. Arai, T. Okuda, A. Tanaka, M. Kotsugi, K. Fukumoto, T. Ohkochi, T. Nakamura, T. Matsushita, T. Muro, M. Oura, Y. Senba, H. Ohashi, A. Kakizaki, C. Mitsumata, and T. Kinoshita, Phys. Rev. B 85, 104418 (2012)

[17] L. Duo, M. Finazzi, and F. Ciccacci (2010). Magnetic Properties of Antiferromagnetic Oxide Materials: Surfaces, Interfaces, and Thin Films, Weinheim: WILEY-VCH Verlag GmbH & Co. KGaA.

[18] L. Liu, T. Moriyama, D. C. Ralph, and R. A. Buhrman, Phys. Rev. Lett. 106, 036601 (2011).

[19] R. Arias, D. L. Mills, Phys. Rev. B 60 7395 (1999)

[20] S. Takei, T. Moriyama, T. Ono, and Y. Tserkovnyak, in preparation.

[21] Y. Tserkovnyak and S. A. Bender, Phys. Rev. B 90, 014428 (2014)

[22] M. Fraune, U. Ruediger, G. Guentherodt, S. Cardoso, and P. Freitas, Appl. Phys. Lett. 77, 3815 (2000)

[23] G. Li, C. W. Leung, C. Shueh, H.-F. Hsu, H.-R. Huang, K.-W. Lin, P. T. Lai, and P. W. T. Pong, Surface & Coatings Technology 228, S437 (2013)

[24] S. I. Kiselev, J. C. Sankey, I. N. Krivorotov, N. C. Emley, R. J. Schoelkopf, R. A. Buhrman, D. C. Ralph, Nature 425, 380 (2003)